# Balanced Receiver Technology Development for the Caltech Submillimeter Observatory

Jacob W. Kooi, Richard A. Chamberlin, Raquel Monje, Brian Force, David Miller, and Tom G. Phillips

*Abstract*—The Caltech Submillimeter Observatory (CSO) is located on top of Mauna Kea, Hawaii, at an altitude of 4.2 km. The existing suite of facility heterodyne receivers covering the submillimeter band is rapidly aging and in need of replacement. To facilitate deep integrations and automated spectral line surveys, a family of remote programmable, synthesized, dual-frequency balanced receivers covering the astronomical important 180–720 GHz atmospheric windows is in an advanced stage of development. Installation of the first set of receivers is expected in the spring of 2012. Dual-frequency observation will be an important mode of operation offered by the new facility instrumentation. Two band observations are accomplished by separating the H and V polarizations of the incoming signal and routing them via folded optics to the appropriate polarization sensitive balanced mixer. Scientifically this observation mode facilitates pointing for the higher receiver band under mediocre weather conditions and a doubling of scientific throughput ($2 \times 4$ GHz) under good weather conditions.

## I. Introduction

ALL the pre-existing facility Superconducting-Insulating-Superconducting (SIS) waveguide receivers at the Caltech Submillimeter Observatory (CSO), Mauna Kea, HI, use waveguide tuners to achieve sensitivities a few times the quantum noise limit. Each of these receivers has played a pioneering role in the submillimeter field. However modern astronomy is demanding more capability in terms of sensitivity, bandwidth, stability, frequency agility, and automation. Although different in detail and configuration, advanced receiver designs are now featured prominently in, for example, the Heterodyne Instrument for the Far-Infrared (HIFI) on the Herschel satellite,[1] ALMA,[2] the Plateau de Bure interferometer (IRAM),[3] the Atacama Pathfinder Experiment (APEX),[4] and the Harvard-Smithsonian Submillimeter Array (SMA)[5].

To upgrade the heterodyne facility instrumentation at the CSO, four tunerless balanced-input waveguide receivers have been constructed to cover the 180–720 GHz frequency range [1]. The new suite of submillimeter receivers will be installed in the Nasmyth focus of the 10.4 m diameter telescope and will soon allow observations in the 230/460 GHz and 345/660 GHz atmospheric windows. The IF bandwidth of the CSO receivers will increase from the current 1 to 4 GHz (though in principle 12 GHz is possible). Balanced configurations were chosen for their inherent local oscillator (LO) spurious tone and amplitude noise cancellation properties, facilitating very stable instrumental baselines, deep integrations, and the use of synthesizer-driven LO chains. It was also judged to be an optimal compromise between scientific merit and finite funding, offering the very stable performance needed to meet the desired science requirements. Unique to the CSO, wide RF bandwidth is favored [2], allowing the same science to be done with fewer instruments. In all the upgrade covers ALMA band 5b-9.

To maximize the RF bandwidth, we explore the use of high-current-density AlN-barrier SIS technology in combination with a broad bandwidth full-height waveguide to thin-film microstrip transition [3]. Compared to $AlO_x$-barriers, advantages of AlN tunnel barriers are a low $\omega RC$ product (increased RF bandwidth) and enhanced chemical robustness. Even if optimal RF bandwidth is not a requirement, a low $\omega RC$ product provides a more homogeneous frequency response and increased tolerance to errors in device fabrication.

To process the required IF bandwidth, the CSO has acquired a fast Fourier transformer spectrometer (FFTS) from Omnisys Instruments, Sweden [4]. This spectrometer facilitates 8 GHz of processing bandwidth with a resolution of 268 kHz/channel, or 3724 channels/GHz. The 8 GHz Omnisys FFTS comes in a 19 inch rack and has two built-in IF processor modules (4–8 GHz each), an embedded controller module, a synchronization module, and power supply. A second rack with an additional 8 GHz of processing bandwidth is provisionally available for special projects.

In this paper we describe the instrument suite and discuss the development of a wide variety of technologies. Particular attention is given to the challenge of providing synthesized LO coverage from 180–720 GHz with minimal latency.

Manuscript received September 06, 2011; revised November 06, 2011; accepted November 19, 2011. This work was supported in part by NSF Grant AST-0838261.
The authors are with the Submillimeter Astronomy and Instrumentation Group, California Institute of Technology, Pasadena, CA 91125 USA (e-mail: kooi@submm.caltech.edu).
Digital Object Identifier 10.1109/TTHZ.2011.2177726

[1][Online]. Available: http://www.sron.nl/divisions/lea/hifi/
[2][Online]. Available: http://www.alma.info/
[3][Online]. Available: http://iram.fr/
[4][Online]. Available: http://www.apex-telescope.org/
[5][Online]. Available: http://sma-www.harvard.edu/

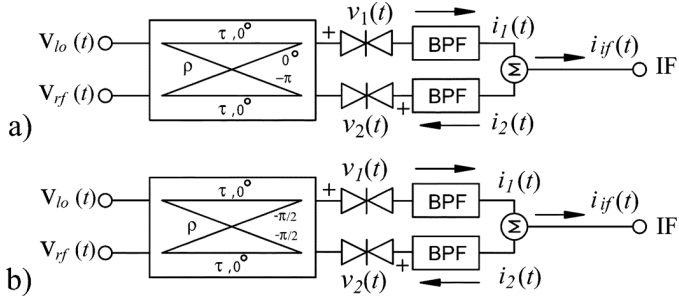

Fig. 1. LO and RF currents in an antipodal biased (single) balanced mixer. In practice, the summing node in the IF can be implemented with an in-phase power combiner [9] or 180° IF hybrid. In the case of the CSO mixers, all IF circuitry is planar and designed using Ansoft's 3-D electromagnetic simulator (HFSS). The band pass filter (BPF) is 3–9 GHz. (a) 180° RF input hybrid. (b) 90° RF input hybrid.

## II. SINGLE-BALANCED MIXER

### A. Introduction

We start with a review of balanced mixer theory, and how this geometry results in a reduction in sensitivity to local oscillator amplitude noise and spurious signals [5]–[7]. In principle, a single balanced mixer can be formed by connecting two reverse biased (SIS) mixers to a 180° or 90° input hybrid, as shown in Fig. 1.

Quantitatively, we can use an exponential to describe the non-linearity of a diode mixer [8]. It should be noted that SIS tunnel diodes have very sharp symmetric I/V curves, and can by their quantum mechanical nature exhibit unity or even positive gain. Using a polynomial series to represent the exponential current $i_1$ through diode 1 we have

$$i_1(t) = a_0 + a_1 v_1(t) + \frac{a_2 v_1(t)^2}{2!} + \frac{a_3 v_1(t)^3}{3!} \ldots = \sum_{n=0}^{\infty} \frac{a_n v_1(t)^n}{n!}. \quad (1)$$

By biasing the second mixer with opposite polarity from mixer 1 we obtain a 180° phase shift so that at the summing node (ignoring device capacitance and the IF bandpass filter)

$$i_{if}(t) = i_1(t) - i_2(t). \quad (2)$$

In this case

$$i_2(t) = b_0 - b_1 v_2(t) + \frac{b_2 v_2(t)^2}{2!} - \frac{b_3 v_2(t)^3}{3!} \ldots$$
$$= \sum_{n=0}^{\infty} (-1)^n \frac{b_n v_2(t)^n}{n!}. \quad (3)$$

Here the terms $a_n$ and $b_n$ represent the mixer conversion gain (magnitude, not power). From (1) and (3) we observe that the term $n = 0$ yields the dc component, $n = 1$ the fundamentals, $n = 2$ the second-order difference and product terms, and $n = 3$ the harmonic and intermodulation products. It should also be noted from (1) and (3) that the product terms decrease by $n!$. Thus we will not consider the fourth or higher order terms.

It is generally known that balanced mixers based on a 180° input hybrid have better LO-RF isolation and improved harmonic intermodulation product suppression than their 90° counterparts. For a 90° hybrid the RF/LO port isolation is found to depend critically on the mixer input reflection coefficient, which for SIS mixers is typically rather poor ($\geq -8$ dB). Unfortunately, 180° hybrids are physically large (Rat-race baluns or waveguide magic Tee's) and difficult to fabricate at frequencies above a few hundred GHz. The analysis presented here evaluates the quadrature single balanced mixer more suitable for submillimeter and terahertz frequencies. The upper and lower sidebands gains are assumed equal, a valid assumption in most instances. In Fig. 1(b) let

$$V_{lo}(t) = V_{lo} \cdot e^{i(m\omega_{lo}t - \frac{\pi}{2})} \quad (4)$$
$$V_{rf}(t) = V_{rf} \cdot e^{i\omega_{rf}t}. \quad (5)$$

An arbitrary LO phase of $-\pi/2$ is used for mathematical simplicity. Defining the power transmission and coupling terms of a hybrid coupler as $\rho^2$ and $\tau^2$ we have for an ideal hybrid $\tau^2 + \rho^2 = 1$ with $\tau^2 = 1/2$. In the case of a non-ideal hybrid we define the power imbalance $G_h = (\rho/\tau)^2$. For an ideal quadrature hybrid the RF voltages $v_1(t)$ and $v_2(t)$ are thus found as

$$v_1(t) = \frac{1}{\sqrt{2}}\left[V_{lo} \cdot e^{i(m\omega_{lo}t - \frac{\pi}{2})} + V_{rf} \cdot e^{i(\omega_{rf}t - \frac{\pi}{2})}\right] \quad (6)$$

$$v_2(t) = \frac{1}{\sqrt{2}}\left[V_{lo} \cdot e^{i(m\omega_{lo}t - \pi)} + V_{rf} \cdot e^{i\omega_{rf}t}\right]. \quad (7)$$

$m$ represents the LO harmonics $1, 2, 3\ldots$. Substituting (6) and (7) into (1) and (3), using $\text{Re}[e^{i\theta}] = \cos(\theta)$ for the second-order term, and summing the IF current as in (2) yields

$$i_{if}(t) = \begin{cases} ia_1\sqrt{2}\left[V_{lo} \cdot e^{i(m\omega_{lo}t)} - V_{rf} \cdot e^{i(\omega_{rf}t)}\right] \\ + \frac{a_2}{2}V_{lo}V_{rf}\cos(|(\omega_{lo} - \omega_{rf})|)t) \\ - \frac{ia_3\sqrt{2}}{4 \cdot 3!}\left[V_{lo}^3 e^{i(3m\omega_{lo}t)} - 3V_{lo}^2 V_{rf} e^{i(2m\omega_{lo}+\omega_{rf})t}\right. \\ \left. + 3V_{lo}V_{rf}^2 e^{i(m\omega_{lo}t + 2\omega_{rf})} - V_{rf}^3 e^{i(3\omega_{rf}t)}\right]. \end{cases} \quad (8)$$

$V_{rf}$ in the first term of (8) is the astronomical signal, which is typically deeply embedded in noise. The second (product) term yields the sum and difference frequencies of the RF and LO signal, i.e., the IF. The third term is found to yield the full complement of third harmonic and intermodulation components. This is unlike the 180° quadrature hybrid [1] with its superior fundamental and intermodulation product suppression capabilities. It explains the popularity of the 180° hybrid at microwave and millimeter wave frequencies. At submillimeter and terahertz frequencies the harmonic and intermodulation products are however severely attenuated by the inherent device capacitance of the mixing element. For this reason, submillimeter or terahertz mixers may be configured with quadrature hybrids, rather than the larger and more complex 180° hybrids.

### B. Amplitude Noise Immunity of a 90° Balanced Mixer

Consider a noise signal $V_n(t) = \Sigma C_k \sin(\omega_k t + \theta_k)$ over all $k$ superimposed on the LO signal. The amplitude(s) $C_k$ and phase(s) $\theta_k$ may be determined from Fourier analyses of $V_n(t)$. In this case the RF voltages at the output of the quadrature hybrid are

$$v_1(t) = [V_{lo}\tau\sin(\omega_{lo}t) + V_n\tau\sin(\omega_n t)] - V_{rf}\rho\cos(\omega_{rf}t) \quad (9)$$

$$v_2(t) = V_{rf}\tau\sin(\omega_{rf}t) - [V_{lo}\rho\cos(\omega_{lo}t) + V_n\rho\cos(\omega_n t)]. \quad (10)$$

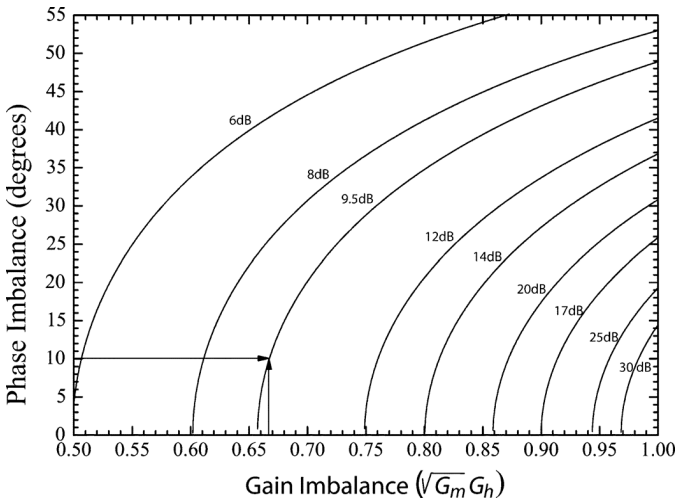

Fig. 2. Amplitude rejection of a balanced mixer relative to an ideal single-ended mixer. Given a realized quadrature hybrid imbalance of 1.2 dB (see Section IV-A), mixer gain imbalance of 1 dB (see Section IV-C), and a differential phase error of $10°$, we can expect an amplitude noise rejection of $\sim 9.5$ dB over the traditional single-ended mixer.

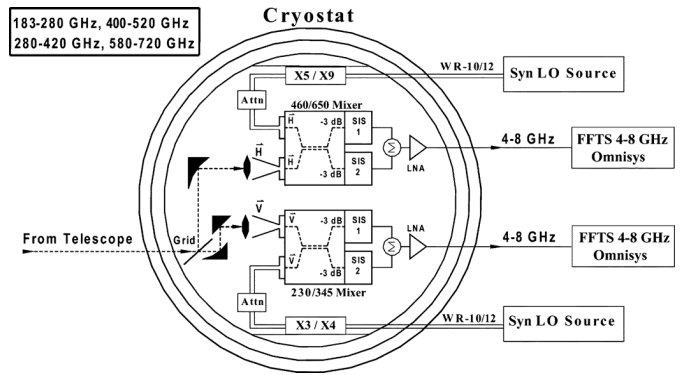

Fig. 3. Cryostat configuration. The 63–105 GHz LO carrier enters the cryostat via an (inner wall) Au-plated stainless steel waveguide (WR-10/12). The submillimeter multipliers [11] are mounted on the 15 K stage of a Precision Cryogenics [12] hybrid cryostat, with LO signals entering the balanced mixers via a cooled fixed tuned (course) attenuator. This is necessary to reduce LO power at the mixer to $\sim 2$ $\mu$W. It also minimizes standing waves between the mixer and multiplier. Each cryostat receives two (orthogonally polarized) beams from the sky, which are routed via a cold wire-grid to the appropriate mixer (see Fig. 4). This technique facilitates dual-frequency (two color) observations, improves observing efficiency, and assists pointing of the high frequency receivers in mediocre weather.

$\omega_n$ represents the frequency components $\Sigma C_k \sin(\omega_k t + \theta_k)$. Substituting $v_1(t)$ and $v_2(t)$ into (1) and (3), and solving the second term of $i_{if}(t)$ (2) for the relevant $V_{lo}(t)V_n(t)$ components while omitting the negligibly small $V_{rf}(t)V_n(t)$ components, we find

$$i_{if}(t) = \frac{1}{2} V_{lo} V_n \cos(\omega_{lo} t - \omega_n t) \cdot [a_2 \tau^2 - b_2 \rho^2 \cos(\Delta\varphi)]. \tag{11}$$

To obtain the noise rejection of the balanced mixer relative to that of a single-ended mixer we divide (11) by the IF noise current in a single-ended mixer $((1)/(2)V_n V_{lo} \cos(\omega_{lo} t - \omega_n t))$. The factor $\cos(\Delta\varphi)$ takes into account the combined phase error of the RF hybrid, device placement, wire bond length, and IF summing node. In this way we obtain

$$\mathrm{NR} = 1 - \frac{b_2}{a_2}\frac{\rho^2}{\tau^2}\cos(\Delta\varphi). \tag{12}$$

Finally, expressing (12) in decibels with the mixer gain imbalance $\sqrt{G_m} = b_2/a_2$, the noise rejection of the balanced mixer relative to a single-end mixer is found to be

$$\mathrm{NR(dB)} = -20 \cdot log[1 - \sqrt{G_m} G_h \cos(\Delta\varphi)]. \tag{13}$$

If, in the configuration of Fig. 1, the mixers are biased symmetric then in the case of a perfectly balanced mixer a doubling of the LO amplitude noise (at the IF summing node) would result [10]

$$\mathrm{NR'(dB)} = -20 \cdot log[1 + \sqrt{G_m} G_h \cos(\Delta\varphi)]. \tag{14}$$

Fig. 2 illustrates the balanced mixer noise reduction as a function of gain and phase imbalance.

### III. CSO NASMYTH FOCUS RECEIVER LAYOUT

The receiver configuration consists of two cryostats, one of which will house the 180–280 GHz/400–520 GHz balanced mixers, the other the 280–420 GHz/580–720 GHz balanced mixers, as shown in Fig. 3.

To supply the needed LO pump power, planar multiplier sources [11] are mounted inside the cryostat and connected to the 15 K stage. This allows for a more compact optical configuration and improves the reliability of the multipliers. We estimate that each SIS junction requires roughly 1/2 $\mu$W of LO pump power ($\alpha = eV_{lo}/h\nu \sim 0.7$ on average). Since two SIS junctions are used as part of the RF tuning design we require $\sim 1.5$–2 $\mu$W of LO power at the mixer LO input port, including waveguide loss in the mixer block.

Given that the cooled multipliers are able to produce ample LO power over the described frequency bands, it is necessary to add attenuation in the LO-mixer path. In practice, this may be accomplished with a directional coupler or fixed tuned (preset) attenuator. A preset attenuator has the advantage of being simple, relatively inexpensive, and manually adjustable at room temperature. The effect of employing a cooled attenuator is similar to the use of a beamsplitter with quasi-optical LO injection; it reduces the multiplier-mixer cavity standing wave, and minimizes additive thermal noise from the local oscillator (see Section V-D). Additional reduction in LO amplitude and spurious noise is provided by the "noise canceling properties" of the balanced mixer as observed from (13).

In Fig. 4 we show the 230 and 460 GHz balanced mixer, LO hardware, and optical components on the 4 K LHe work surface.

### IV. MIXER BLOCK HARDWARE

#### A. Quadrature-Hybrid Waveguide Coupler

The E-plane split-block quadrature hybrids are based on earlier work by Claude and Cunningham *et al.* [13] for the Atacama Large Millimeter Array (ALMA),[2] but modified to maximize RF bandwidth and optimize machinability (Fig. 5). We utilized HFSS [14] in the design and optimization of the wideband quadrature-hybrid couplers. The design parameters are compiled in Table I, and the predicted performance for the 280–420 GHz hybrid shown in Fig. 6.

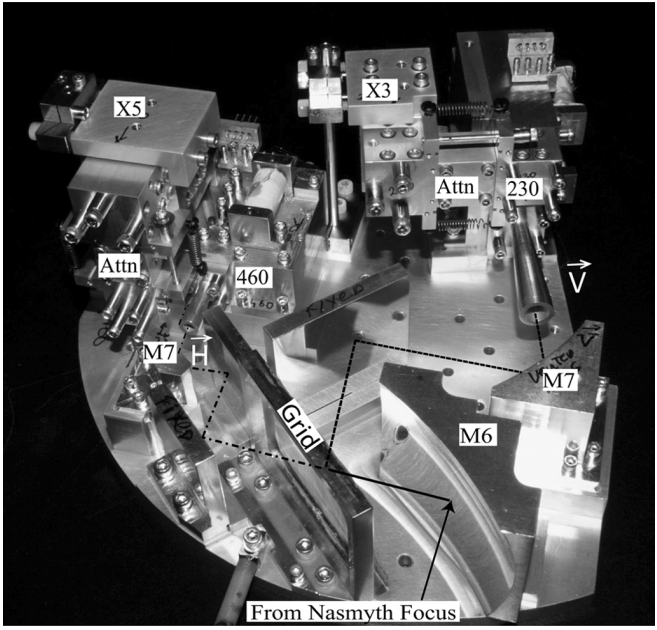

Fig. 4. 230/460 GHz FPU with associated balanced mixer blocks, multiplier hardware, and optics.

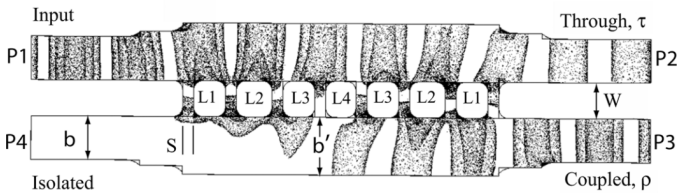

Fig. 5. Electric field distribution in the quadrature waveguide coupler depicting a $90\pm1.5°$ phase difference and 3 dB power split between output ports 2 and 3. $S$ sets the coupling, and $W$ the center frequency. $L1$–$L4$ are optimized to minimize the input return loss. $a$ and $b$ are the hybrid waveguide dimensions. See Table I for dimensions.

TABLE I
QUADRATURE HYBRID COUPLER PARAMETERS

| $\nu_{ct}$ | a | b | b' | S | W | L1 | L2 | L3 | L4 |
|---|---|---|---|---|---|---|---|---|---|
| GHz | μm | μm | μm | μm | μm | μm | μm | μm | μm |
| 230 | 1040 | 414 | 568 | 114 | 305 | 318 | 356 | 318 | 292 |
| 345 | 680 | 270 | 371 | 74 | 201 | 206 | 231 | 206 | 190 |
| 460 | 528 | 211 | 287 | 86 | 147 | 140 | 168 | 142 |  |
| 650 | 363 | 145 | 198 | 74 | 102 | 99 | 117 |  |  |

In order to achieve the required bandwidth, phase, and coupling, it was necessary to maximize the number of branches $(n)$. Based on the computer simulation results it was determined that

$$n * S \simeq \frac{\lambda_g}{2} \ \& \ W \simeq \frac{\lambda_o}{4} \quad (15)$$

where $n$ denotes the number of coupling sections, $S$ the branch line width, $W$ the hybrid waveguide spacing, $\lambda_g$ the guide wavelength, and $\lambda_o$ the wavelength of free space. Here $S$ sets the coupling imbalance $G_h = (\rho/\tau)^2$, and $W$ the center frequency of the coupler. After consulting with the manufacturer [15] it was determined to fix the minimum width of the branch line to 75 μm. To now allow for a maximum number of branches the width of the waveguide inside the coupler was increased by 32.5% ($b'$). This effectively decreases the Electric field density in the hybrid, thus allowing for either an increase in the branchline width and/or an increase in the number of coupling sections $(n)$. Increasing the waveguide width any further excites the $TE_{01}$ mode, thereby degrading the high frequency performance of the coupler.

### B. Integrated IF and Wilkinson In-Phase Summing Node

In a mixer configuration, the active device is typically terminated into a desired IF load impedance, the bias lines EMI-filtered and injected via a bias Tee, and the IF output dc-isolated (see Fig. 7). The balanced mixer has the additional constraint that the IF signals need to be combined either in phase, or 180° out of phase, putting tight limits on the allowed phase error ($<5°$). Since in our application the SIS junctions will be biased antisymmetric (see Fig. 1) we conveniently combine the bias-Tees, electrical isolation of the IF port, band pass filters, IF matching networks, and an in-phase Wilkinson power combiner [9] on a single planar circuit. The 100 Ω balancing resistor of the Wilkinson power combiner [see Fig. 7(b)] is a 1% laser trimmed thinfilm NiCr resistor, lithographically deposited on a

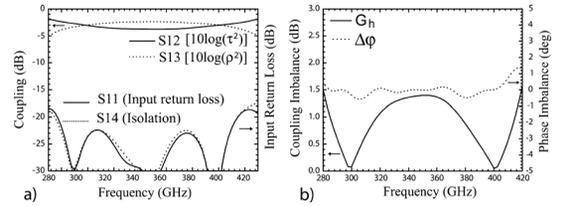

Fig. 6. a) Predicted coupling performance of the 280–420 GHz 90° hybrid. $\tau$ and $\rho$ are the quadrature hybrid transmission and coupling coefficients. b) Phase and power imbalance $(10 \cdot \log[(\rho/\tau)^2])$ of the hybrid coupler. Based on computer simulations we expect 0.4–0.5 dB of cold absorption loss in the coupler.

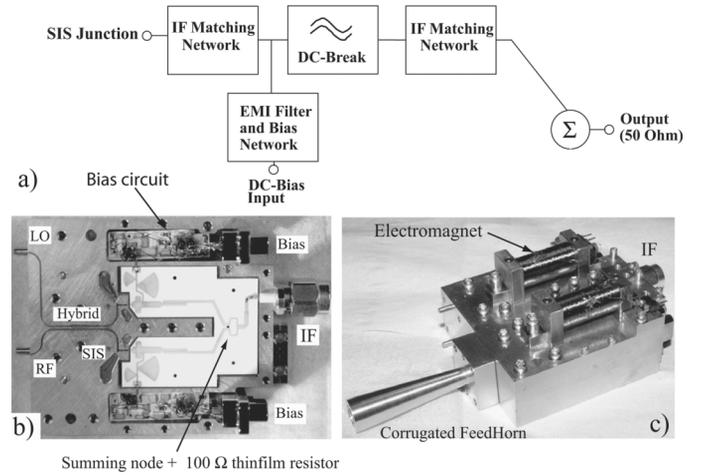

Fig. 7. a) Balanced mixer block layout. b) The IF board is entirely planar (alumina), and combines the IF match, dc-break, bias Tee, EMI filter, and Wilkinson in-phase power combiner. The E-field component of the incoming signal is horizontally polarized along the waveguide split. The calculated RF path cold waveguide loss ranges from 0.16 dB at 230 GHz to 0.31 dB at 0 650 GHz. c) Josephson noise suppression in the SIS tunnel junctions is accomplished by two independent electromagnets.

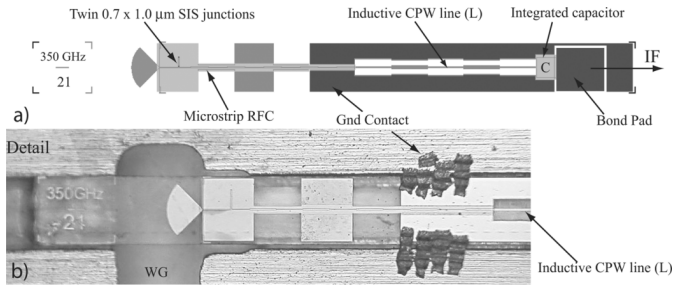

Fig. 8. a) Twin-SIS junction 350 GHz chip layout. The radial probe waveguide antenna is visible on the left-hand side. The IF is taken out via a microstrip RF choke (on 300 nm SiO, $\epsilon_r = 5.6$) which connects to a high impedance CPW transmission line (inductive) and shunt capacitor. This LC mechanism provides a $\pi$ tuning network with the combined capacitance of the probe, twin junction RF tuning structure, and microstrip RF choke. The IF passband has been optimized to cover 1–13 GHz. b) Example of a 345 GHz chip mounted in a waveguide.

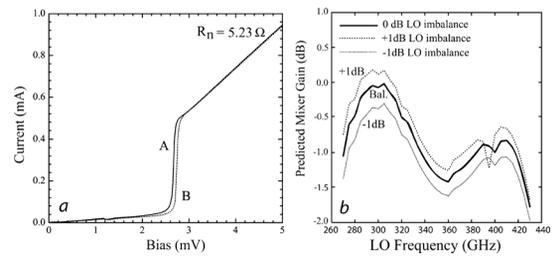

Fig. 9. (a) Measured unpumped I/V curves of B030926 [18]. The difference in gap voltage does not significantly impact the mixer noise performance. (b) The effect on the 345 GHz mixer gain by unbalancing the LO signal power level by $\pm 1$ dB.

635 $\mu$m thick Alumina ($\epsilon_r = 9.8$) circuit board [16]. This compact choice conveniently avoids the use of a physically larger (commercial) 180° hybrid.

The IF bandpass filter is comprised of a set of parallel coupled suspended microstrip lines [17]. For this filter to work, the ground plane directly underneath the filter has been removed, and the IF board positioned on top of a machined cutout (resonant cavity). There are several discontinuities in this structure. When combined, they form the bandpass filter poles. The advantages are; simplicity of design (only one lithography step), accurate knowledge of the phase, and reliability. The disadvantage is possibly its size, $\lambda_g/4$ ($\sim 6$ mm at 6 GHz).

### C. High Current Density SIS Junctions With Integrated IF Matching

To facilitate the CSO heterodyne upgrade a suite of high-current-density AlN-barrier niobium SIS junctions (four bands) have been fabricated by JPL [1]. These devices have the advantage of increasing the mixer instantaneous RF bandwidth while minimizing absorption loss in the mixer normal or superconducting thinfilm front-end RF matching network.

The new SIS tunnel junctions of Fig. 8 all share the same 50 $\mu$m thick quartz wafer (B030926 [18]). This has as benefit that a successful wafer run contains all the mixer chips needed for the 180–720 GHz facility receiver upgrade. The junction designs employ twin-SIS junctions with a $R_n A$ product of 7.6 $\Omega \cdot \mu m^2$ ($J_c = 25$ kA/cm$^2$ current density). Supermix [19], a flexible software library for high-frequency superconducting circuit simulation, was used in the design process.

In Fig. 9(a) we show the measured 345 GHz balanced mixer I/V curves (2). In general, the junction characteristics are reasonably well matched, with slight variations in the definition of the energy gap and device area. The depicted devices were selected on merit of matching I/V curves, e.g., normal state resistance ($R_n$), leakage current at 2 mV bias, gap voltage ($V_{gap} = 2\Delta/e$), and sharpness of the energy gap (d(I/V)/dV). The measured junction normal state resistances of Fig. 9(a) are 3.79% ("A") and 3.52% ("B") below the theoretical value of 5.43 $\Omega$.

A limitation of the balanced hybrid design is that the LO power as a function of frequency is not equally split between the two junctions [see Fig. 6(b)]. This situation for a $\pm 1$ dB LO pump imbalance is shown in Fig. 9(b). From analysis we conclude that gain imbalance due to device characteristics and LO power imbalance is not expected to significantly affect the overall balanced mixer performance. This is important since it means that the individual SIS junctions may be biased at similar, but opposite polarity. The simulation results are derived from harmonic balanced superconducting SIS mixer simulations [19] in combination with extensive Sonnet [20] and HFSS [14] analysis of the RF and IF mixer circuitry.

### V. Synthesized Local Oscillator (186–720 GHz)

The CSO suite of balanced receivers will employ a dual-synthesizer LO configuration operating at a baseband frequency of 20–35 GHz (see Fig. 10). This setup facilitates remote and automated observations, frequency agile performance, and ease of operation. The commercial synthesizers [21] connect via 2.5 m length low loss coaxial cables [22] to a pair of magnetically shielded signal conditioning "mu-Boxes". Each mu-Box contains a medium power amplifier (Psat = +25 dBm) [23], a tunable Yttrium-Iron-Garnet (YIG) 4-pole bandpass filter [24] for removal of low level spurious content, a 20 dB directional coupler and zero bias detector diode for the purpose of signal monitoring/calibration, a 3 dB power splitter, and two Ditom $K_a$-band isolators [25]. The input signal to the YIG is approximately +22 dBm. The isolated output ports ($\sim +11$ dBm) route the filtered carrier signal via 30 cm and 1.4 m coaxial lines to a second signal conditioning box, known as the "mm-Wave" box.

Like the mu-Box, the mm-Wave signal conditioning box also contains medium power amplifiers [23]. These also are run into saturation (+25 dBm) thereby reducing amplitude noise on the carrier. Fourier harmonics from the resulting clipped sinusoidal waveform are removed by means of a 35 GHz 17-pole low pass filter (LPF) [26]. The measured in-band signal loss of the lowpass filter at 35 GHz is $\sim 1.75$ dB. At 40 GHz the attenuation has increased to $\sim 23$ dB. In large part due to frequency dependent variations in the saturated output power of the mm-wave medium power amplifiers, the available signal level to drive broad bandwidth passive triplers [27] ranges from +20 to +22 dBm. The output of the passive triplers is either WR-12 or WR-10 waveguide (TE$_{10}$ mode), depending on the frequency band. Given the 17 dB conversion loss of the triplers and 1 dB waveguide loss, this translates into a (measured) input signal level at the WR-11 power amplifiers (see Section V-B) of 1–2 mW.

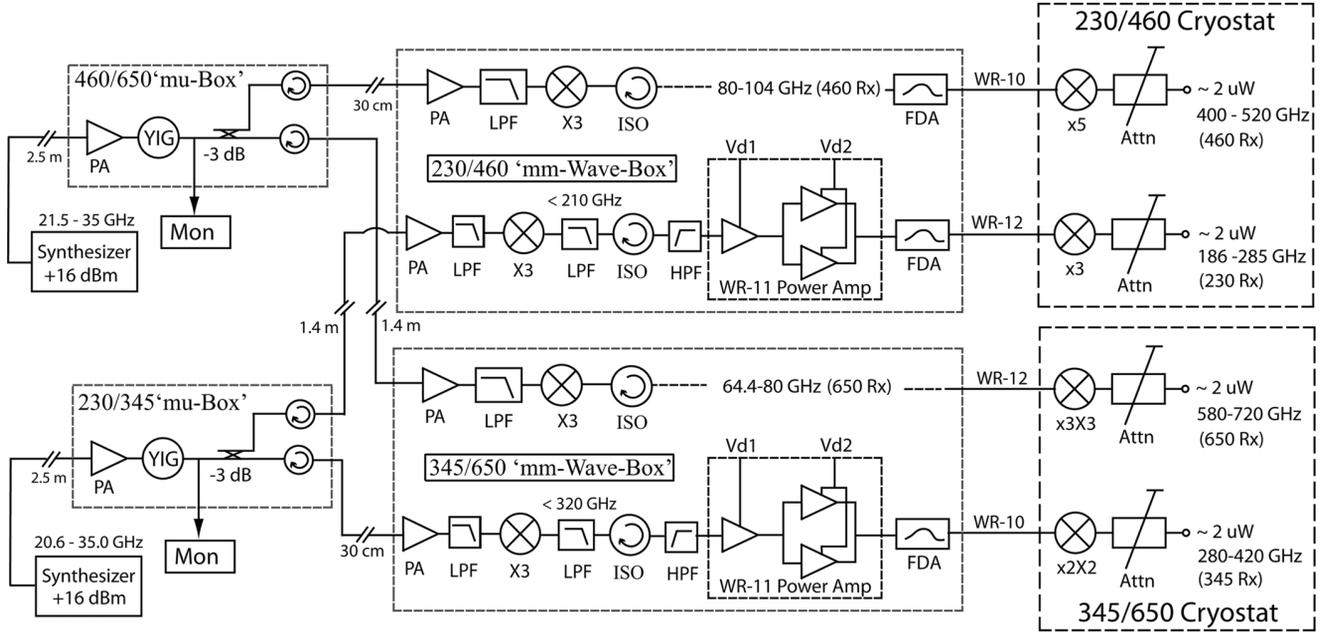

Fig. 10. CSO dual-frequency synthesized local oscillator layout. At the input of the mu-Box the baseband frequency of 20–35 GHz ($K_a$-band) is amplified and drives the medium power amplifier into saturation. The LO signal is filtered by the YIG to remove low level spurious and harmonic content, passively multiplied (X3) to 62–105 GHz, band pass filtered to avoid unwanted second and fourth order harmonics, once again amplified (WR-11 waveguide power amplifiers), signal conditioned (FDA), and finally injected into the cryostat where the carrier signal is multiplied up to the final submillimeter frequency (186–720 GHz) and injected into the balanced mixers via a cooled attenuator. Spectral line observations below 186 GHz will need to be in the mixer lower side band.

TABLE II
MULTIPLICATION FACTORS OF THE CSO SYNTHESIZED LO

| Frequency | PMW [32] | VDI [16] | Total Multiplication |
|---|---|---|---|
| 186-280 GHz | ×3 | ×3 | ×9 |
| 280-420 GHz | ×3 | ×2×2 | ×12 |
| 400-520 GHz | ×3 | ×5 | ×15 |
| 580-720 GHz | ×3 | ×3×3 | ×27 |

Following the WR-11 balanced power amplifier there is a "Frequency Dependent Attenuator" (see Section V-C) which is designed to provide an optimum (safe) drive level for the VDI [11] passive multipliers (see Table II) at the 15 K work surface of the cryostat. Finally, to inject the 186–720 GHz submillimeter LO signal into the balanced mixers (see also Fig. 3), while providing a suitable level of attenuation and thermal break, a cooled waveguide attenuator is employed (see Section V-D).

At frequencies below 210 and 320 GHz there is the possibility that harmonics at the high end of the frequency band will be amplified by the WR-11 power amplifier, and thus be incident along with the intended carrier frequency on the final multipliers. To eliminate this possibility a set of waveguide lowpass filters has been designed [14] and procured [28]. The mm-Wave boxes are designed to facilitate insertion of the lowpass filters in the LO path as needed. Additionally, to choke 2nd order harmonic content from the passive tripler a 2.5 cm "split-block" section WR-9.5 and WR-8.5 waveguide is employed at the output of the 230/345 band isolators (see Fig. 10).

### A. YIG Filter

We have measured a 500 GHz SIS heterodyne receiver driven by a commercial synthesizer as an LO source [21]. Without a YIG tracking filter [24] at the output of the synthesizer we observed significant spurious content in the IF spectrum from the SIS receiver. With a YIG tracking filter the receiver IF spectrum was free of spurious tones. This result is understood in that the selected YIG filter has a specified out of band spurious signal rejection of 60–80 dB, providing $\geq$ 40 dB of additional rejection over a commercial synthesizer. The configuration of our synthesized LO system is shown in Fig. 10.

Additional testing on the YIG tracking filters showed that the filter passband tends to drift with time (over days) and that the performance was improved if the YIG case was temperature stabilized with an external heater [29]. Further, since the YIG requires a strong magnetic field for tuning it had to be packaged in a mu-metal enclosure ("mu-Box") ($\sim$ 75% nickel, 15% iron, copper, molybdenum) so that it would not interfere with the SIS mixer which is in proximity. The YIG mu-metal enclosures also contains: an SPI programming interface to the receiver control computer; power detection for monitoring and computer closed loop control of the YIG programmed center frequency; and, temperature control, monitoring, and over-temperature safety shutdown.

Important science considerations to the operation of the synthesized LO, and in particular the YIG tracking filter, are the rate at which the YIG tuning parameters (due to drift/hysteresis) need to be updated and the settling time after a retune. In particular, having optimal YIG tuning parameters (the YIG filter is of the open-loop type [30]) is important in keeping the YIG bandpass centered on the synthesizer carrier frequency.

From our extensive testing [31] we formed the following conclusions.

1) The YIG filter short term tuning parameter drift and residual tuning error ($\Delta_f$) is smaller with the case tem-

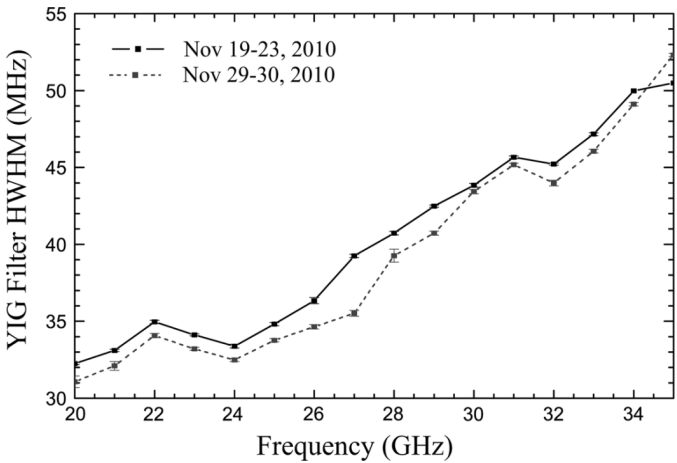

Fig. 11. YIG bandpass HWHM versus frequency taken at two time periods; Nov. 19–23 and Nov. 29–30, 2010. The standard deviation on the properly tuned (Nov. 19–23) data set is ± 0.10 MHz, whereas the standard deviation on the drifted measurement (Nov. 29–30) was ± 0.18 MHz as indicated by the error bars. The YIG case temperature was 80 °C.

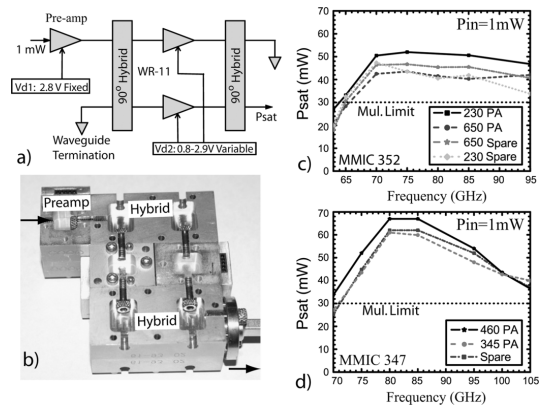

Fig. 12. a) Pre-amp and balanced power amplifier setup. The pre-amp is fixed biased at ∼ 2.8 V and drives two variable bias gain modules in a balanced configuration. b) Assembled power amplifier. The measured RF loss in the quadrature hybrids is ∼ 1 dB. c) Measured output power of the MMIC 352 configuration. This chain is used to drive the 230 and 650 GHz submillimeter multipliers. d) Measured output power of the MMIC 347 configuration which is used to drive the 345 and 460 GHz submillimeter multipliers. See also Table II. The input power level requirement is ≥ 1 mW. Note the individual variation.

perature warm, near 80 °C, than cool. This result is not consistent with the manufacturer's recommendation that 0 °C < case temperature < 65 °C.

2) The YIG filter actual center frequency ($f_a$) settles fastest with the case heater at ∼77 °C and the internal YIG heater ON, as opposed to with the internal YIG heater OFF.

3) There may be substantial drift in the YIG tuning parameters on the time scale of days. Micro-Lambda confirmed that the observed drift is consistent with their experience. Thus, the YIG tuning curve should be checked at least daily and may require a "peak-up" tuning algorithm.

4) A wait time of ∼ 2.5 min should be enforced for large frequency jumps, defined as ≥ 5 GHz to allow the YIG to settle before observing.

Fig. 11 depicts the measured YIG filter half width at half max (HWHM).

### B. 63–105 GHz Waveguide Power Amplifiers

The output signals of the (passive) millimeter wave triplers [27] have to be amplified before they can be routed into the cryostat and drive the final submillimeter multipliers [11] (see Fig. 10). This amplification is achieved with custom designed balanced power amplifiers (PAs). Every PA consist of three WR-11 gain modules, each of which houses a medium power monolithic microwave integrated circuit (MMIC) [32] originally developed for HIFI [33], the high resolution instrument on Herschel [34].

The Coplanar Waveguide (CPW) chips use 0.1 $\mu$m AlGaAs/InGaAs/GaAs pseudomorphic T-gate power HEMT MMIC technology on 50 $\mu$m thick GaAs substrate. The MMICs cover the 70–100 GHz (MMIC 352) and 80–115 GHz (MMIC 347) frequency range, and have output power levels of 25–50 mW. For the HIFI gain modules, WR-10 and WR-8 waveguide were employed with $TE_{10}$ mode frequency limits of approximately 70 and 88 GHz. Inspection of on-wafer S-parameter measurements reveals [32] that the CPW-MMIC driver chips are capable of an extended frequency range, albeit with slightly reduced performance. To accommodate the CSO 63–105 GHz synthesized LO requirement, WR-11 waveguide gain modules were designed, with twenty fabricated, tested, and assembled into the PA configuration of Fig. 12. An advantage of using WR-11 based power amplifiers is that they may be interfaced to WR-10 and WR-12 waveguide with a return loss ≤ −20 dB. The measured performance of the "low" frequency (MMIC 352) and "high" frequency (MMIC 347) PAs is shown in Fig. 12(c) and (d).

The use of balanced amplifiers has the advantage that RF power is combined, and reflected power terminated into a (internal) load [35]. This is particularly important as the input- and output return loss (IRL/ORL) of the MMIC chips can be as high as −4 dB at the (extended) band edges. The balanced configuration has a measured ORL of ≤ −17 dB thereby minimizing standing waves at the output port. To minimize reflections of the single-ended input pre-amp we employ full-waveguide band WR-12 and WR-10 isolators [36]. With this configuration, and the pre-amp module fixed biased at 2.8 V, the output of the balanced amplifiers was measured to be in saturation over the entire range of usable drain voltages (0.8–2.9 V). The latter being very important in minimizing LO AM-carrier noise.

### C. Frequency Dependent Attenuators

The balanced power amplifiers of Section V-B, though extended in RF bandwidth, have significant variation in saturated output power (∼ 4 dB). This characteristic is problematic when driving submillimeter multipliers with typical input power level requirements of ≤ 30 mW. To constrain the available RF power to safe levels [see Fig. 12(c) and (d)] it was decided to provide hardware limits by means of "Frequency Dependent Attenuators" or FDAs. This is opposed to software limits with safety tables, as is the case with HIFI [33]. The result is shown in Fig. 13.

Conceptually the FDA may be thought of as a flute with frequency selective (tuned) branches; six in this case [see Fig. 13(b)]. Like a flute, each branch resonates at a particular frequency, the combined effect giving the "sound" or passband. The terminating waveguide loads are based on an ALMA design [35]. The power transmission was designed such that $P_{sat} \cdot |S21|^2_{FDA} \sim 30$ mW. The only exception is the 650 GHz integrated ×3×3 multiplier (see Table II) which due to its limited frequency range and large power handling capability (80 mW) does not require a FDA. The (measured) signal conditioned RF power including FDA is shown in Fig. 13(d).

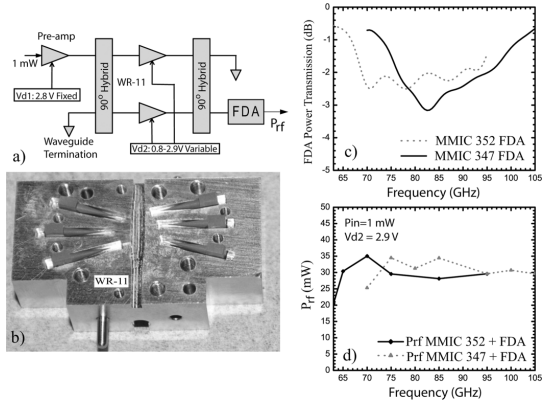

Fig. 13. a) Pre-amp and balanced power amplifier setup including a "frequency dependent Attenuator" (FDA). b) Photograph of the (E-plane) split block showing the six frequency tuned branches with waveguide absorbers. c) Calculated power transmission through the FDA. d) Measured output power, $P_{rf}$. The measured output return loss is $< -18$ dB.

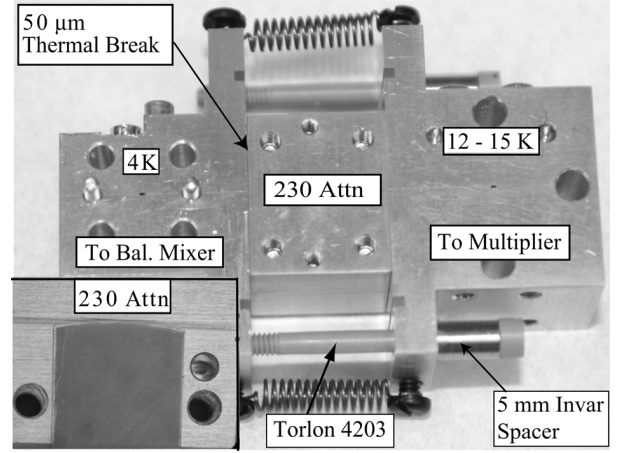

Fig. 14. 230 GHz "cooled" waveguide attenuator, thermal break, and flange adaptor. The overall size of the unit is 4.9 cm × 1.9 cm × 1.4 cm. The attenuator is formed by manually inserting a 50 $\mu$m resistive card (inset) into the split-block waveguide. See text for details.

### D. LO Injection

As shown in Fig. 4, the final LO multipliers [11] are mounted to the cryostat LHe temperature cold work surface while thermally strapped to an intermediate 15 K cold stage. This is done for stability reasons (curtail mechanical modulation of the LO-mixer standing wave [37]–[39]) and to minimize thermal heat loading on the LHe reservoir (multipliers have a low conversion efficiency and most of the RF power is converted into heat). For the CSO Nasmyth receivers, injection of the LO signals quasi-optically from outside the cryostat was never an option as this required new and larger cryostats, both of which were beyond the scope of the upgrade effort. Since the LO signal is injected via waveguide (see Fig. 10) a vacuum block at the cryostat entrance is needed. For this we use 0.27 mm thick Mica ($\epsilon_r = 2.54 - 2.58$) [40], building on ALMA Band 9 [41] development. A dc-break is also provided at this junction to avoid ground loops.

To connect the LO signal from the (final) multiplier output to the input port of the balanced mixer a number of technical criteria have to be satisfied: First, there has to be an in-line (course) attenuator to set the LO power to an appropriate level. Second, there needs to be a thermal break between the 15 K multiplier and 4 K mixer. Third, flange adaptors are needed on both the multiplier and balanced mixer ports. And finally, the whole configuration has to be compact to minimize RF loss and fit the limited workspace. Fig. 14 shows the final arrangement.

The attenuator is formed by inserting a 50 $\mu$m "resistive" card, its surface along the electric field, into a centrally located non-radiating slit in the broad wall of the waveguide. The high surface resistance aluminized Mylar card consists of a nm thick metal film with a penetration depth less than a skin depth (local limit). To measure the RF output power we use an "Erickson" style calorimeter [42], [11] in automated synchronous detection mode by switching the balanced power amplifier (see Section V-B) ON/OFF at an appropriate rate ($\sim 30$ s). The calorimeter Allan variance response time [38] was improved by mounting it to a copper baseplate with neoprene insulation throughout, providing a rms noise level of $\sim 0.1$ $\mu$W.

The thermal break is accomplished via a 50 $\mu$m airgap, formed by eight $1 \times 1$ mm Kapton spacers, located at the perimeter of the attenuator and mixer block interface. Torlon 4203 #4 screws in combination with insulated SS springs hold the blocks together. Torlon was chosen for its very low thermal conductivity at 4 K and high strength. The two SS springs in Fig. 14 provide a constant tension and long thermal path. Specially designed 5 mm invar spacers compensate for the difference in thermal contraction between the gold plated brass blocks and Torlon screws. To minimize RF loss in the thermal break, a quarter wave (circular) RF choke is incorporated in the waveguide flange on both sides of the break. To verify the integrity of the thermal break an Ohm meter may be used. Fig. 15 shows the derived waveguide attenuation between 186–720 GHz at room temperature and the measured LO power. The atmosphere is provided for reference. It should be noted that the multiplier efficiency is expected to improve upon cooling by approximately 25%–40%.

### E. Multi-Layer Vacuum Window and IR Block

Because of its high transparency into the submillimeter, relatively low dielectric constant ($\epsilon_r = 3.8$), and opaqueness to helium and other gasses, Z-cut quartz is commonly used for vacuum windows and IR filter applications. However, without anti-reflection coating the reflected power loss is substantial ($\sim 10\%$). In the case of quartz, only Teflon has the theoretically ideal dielectric constant and low loss tangent to serve as anti-reflection material. Unfortunately, adhesion of Teflon to quartz is difficult and costly. Other materials such as silicon ($\epsilon_r = 11.8$) with Parylene-c [43] as anti-reflection coating have been investigated. However the dielectric constant of Parylene-c ($\epsilon_r = 2.62$) and material absorption loss are not ideal, vacuum deposition specialized, and signal bandwidth limited due to the high dielectric constant of silicon.

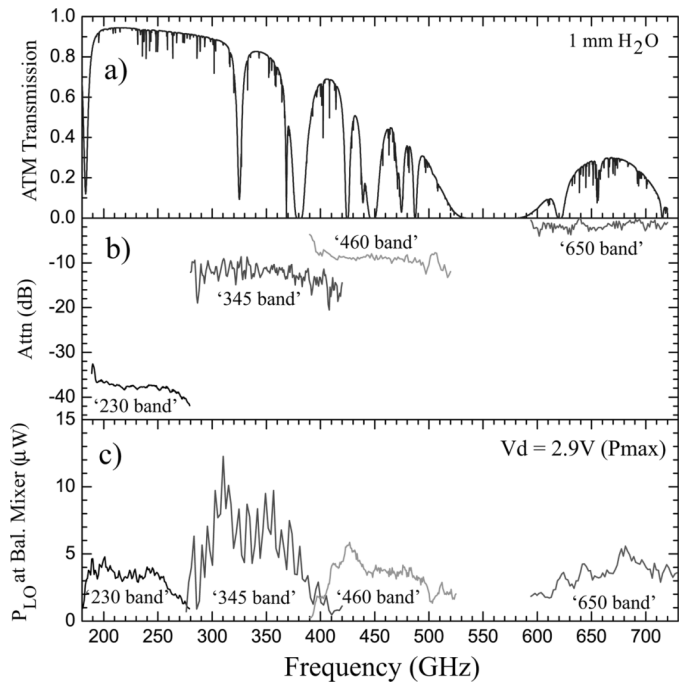

Fig. 15. a) Atmospheric transmission over Mauna Kea, HI, for 1 mm of precipitable water vapor. b) Derived attenuation for the four waveguide bands. c) Measured available LO power at the input of the balanced mixer LO port (300 K). The standing wave in the 345 GHz LO power response results from interaction between the $\times 2 \times 2$ multiplier stages (see Table II). This is expected to improve by biasing the diodes in constant current (CI) mode.

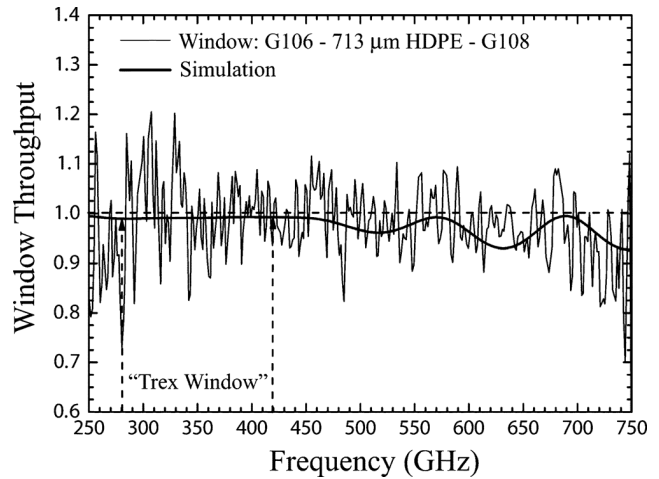

Fig. 16. Measured and modeled 713 $\mu$m HDPE window with G106 (150 $\mu$m) and G108 (200 $\mu$m) Zitex AR coating. The in-band reflection loss is on the order of 2%. This window is used on the "Technology Demonstration Receiver" of Section VI.

TABLE III
VACUUM WINDOW PROPERTIES

| Frequency Band (GHz) | Material Thickness ($\mu$m) | | | | |
|---|---|---|---|---|---|
| | $t_1$ | $t_2$ | $t_3$ | $t_4$ | $t_5$ |
| 280-420 | 152 | 50 | 713 | 50 | 203 |
| 180-280 & 400-520 | 152 | 50 | 726 | 50 | 203 |
| 280-420 & 580-750 | 102 | 50 | 756 | 50 | 152 |
| Dielectric Constant ($\varepsilon_r$) | 1.44 | 2.29 | 2.35 | 2.29 | 1.44 |

An interesting alternative to quartz is high-density polyethylene (HDPE). This material is used by many research groups to make lenses, infrared-blocking filters, and vacuum pressure windows. HDPE has good transmission well into the far infrared, good to excellent blockage of helium and other atmospheric gasses, a low dielectric constant (2.35), and low loss tangent ($\tan d\delta \sim 0.0005$), resulting in $\sim$4.5% mean power reflection loss. To further reduce this loss we have successfully anti-reflection coated HDPE with Zitex [44] (see Fig. 16). And to maximize the RF bandwidth while minimizing reflection it was found that the use of Zitex with different sheet thickness on either side of the HDPE window is preferred.

Zitex, being 50% porous Teflon, has a melting point of $\sim$327 °C, significantly higher than that of HDPE (130 °C–137 °C). Using a 50 $\mu$m layer of low-density Polyethylene (LDPE) [45] with a melting point $\sim$10 °C below that of HDPE as a glue, we have under pressure (1/4 ton) and heat (125 °C) fused a "sandwich" of Zitex-LDPE-HDPE-LDPE-Zitex ($t_1$, $t_2$, $t_3$, $t_4$, $t_5$ in Table III). There is a 7%–10% shrinkage in the HDPE thickness due to the applied pressure and temperature.

As a final note, ALMA [46] has developed excellent five-layer quartz vacuum windows (Zitex-HDPE-Quartz-HDPE-Zitex), albeit at a significant complexity and cost. The performance of the described Zitex-LDPE-HDPE-LDPE-Zitex "sandwich" is very competitive with these excellent multi-layer windows. Because the thinnest sheet of Zitex is 100 $\mu$m (G104), the application of the multi-layer Zitex-LDPE-HDPE-LDPE-Zitex technique is limited to $\sim$ 750 GHz.

## VI. SENSITIVITY

A single-ended "Technology Demonstration Receiver" (Trex) covering the important 275–425 GHz atmospheric windows was installed at the CSO in 2007 [2]. Trex is in active use and offers an unprecedented 43% RF bandwidth, nearly 50% wider than the ALMA band 7 275–373 GHz specification [47]. The Trex instrument has proven itself to be an extremely useful testbed for the many new and exciting technologies outlined in this paper.

In Fig. 17 we show the simulated balanced receiver and mixer noise temperature from 180–720 GHz in 4 waveguide bands. To derive a realistic estimate for the balanced receiver noise, we used the measured optics losses of the existing CSO receivers minus LO thermal noise. Superimposed in the plot are the measured results of the single-ended "Technology Demonstration Receiver". From the discussions it is evident that the primary advantage of the balanced receiver is not sensitivity, but rather LO amplitude and spurious noise suppression, stable operation (deep integrations), high quality baselines, and efficient use of local oscillator power [5].

## VII. CONCLUSION

The facility receivers of the CSO are being replaced with fully synthesized, dual-frequency, tunerless, 4–8 GHz IF bandwidth state-of-the-art versions. At their heart, the new Nasmyth heterodyne instrumentation will consist of four balanced mixers designed to cover the 180–720 GHz atmospheric frequency range (ALMA B5b-B9). Design and fabrication of the many individual components, e.g., low noise amplifiers, SIS junctions, mixer blocks, corrugated feedhorns, optics, synthesized LO, etc., is now complete with assembly and characterization in full swing.

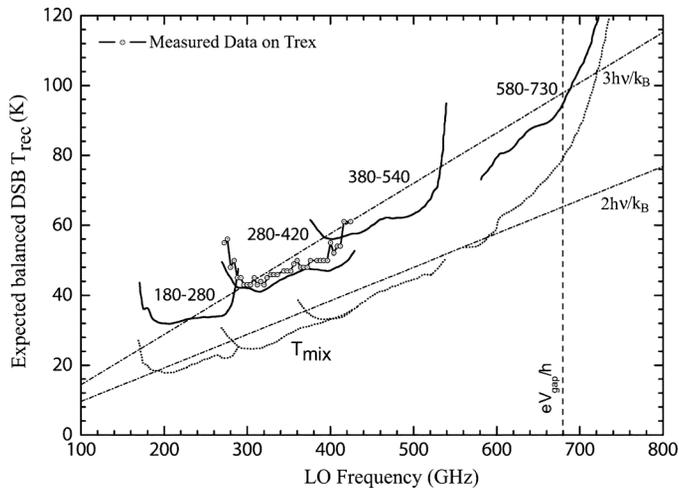

Fig. 17. Estimated double sideband receiver noise temperatures for the new suite of balanced mixers. The noise estimate was calculated by Supermix [19], and includes a realistic optics and IF model. The LO noise contribution should be negligible owing to the noise rejection properties of the balanced mixers (calculated to be $\sim$ 9.5 dB), and the cooled LO/attenuator. The balanced mixer noise follows the $2h\nu/k_B$ line and is useful to estimate receiver temperatures for different IF and optics configurations.

To facilitate automated tuning procedures, remote observations, spectral line surveys, frequency agility, ease of operation, and enhanced scientific throughput (scripting), the receivers and local oscillators will be under full synthesizer and computer control. Furthermore, to accommodate the 2×4 GHz IF bandwidth, the CSO has recently installed an 8 GHz Fast Fourier Transformer Spectrometer (FFTS) with a channel resolution of 268 kHz.

ACKNOWLEDGMENT

The authors would like to thank M. Gould of Zen Machine and Scientific Instruments for his mechanical advice and machining expertise, J. Pierson of the Jet Propulsion Laboratory for his assistance with the medium power amplifiers modules, J. Parker for assembly of the many LO related components, K. Cooper for setting up the data acquisition network, and J. Groseth for help with the Cryogenics and laboratory work. We also wish to thank Pat Nelson for rewiring of the Cryostat and the CSO day-crew for their logistic support over the years. They would also like to thank Prof. J. Zmuidzinas of Caltech and the former CASIMIR program for providing the $K_a$-band synthesizers and Fast Fourier Transform Spectrometers (FFTS).

REFERENCES

[1] J. W. Kooi, A. Kovács, B. Bumble, G. Chattopadhyay, M. L. Edgar, S. Kaye, R. LeDuc, J. Zmuidzinas, and T. G. Phillips, "Heterodyne instrumentation upgrade at the Caltech submillimeter observatory," in *Proc. SPIE*, 2004, pp. 332–348.
[2] J. W. Kooi, A. Kovács, M. C. Sumner, G. Chattopadhyay, R. Ceria, D. Miller, B. Bumble, R. LeDuc, J. A. Stern, and T. G. Phillips, "A 275–425 GHz tunerless waveguide receiver based on AlN SIS technology," *IEEE Trans. Microw. Theory Tech.*, vol. 55, no. 10, pp. 2086–2096, Oct. 2007.
[3] J. W. Kooi, G. Chattopadhyay, S. Withington, F. Rice, J. Zmuidzinas, C. K. Walker, and G. Yassin, "A full-height waveguide to thin-film microstrip transition with exceptional RF bandwidth and coupling efficiency," *Int. J. Infrared Millim. Waves*, vol. 24, no. 3, pp. 261–284, Sep. 2003.
[4] Omnisys Instruments AB, Västra Frölunda, Sweden, "Omnisys homepage," 2009. [Online]. Available: http://www.omnisys.se
[5] P. D. Strum, "Crystal checker for balanced mixers," *IEEE Trans. Microw. Theory Techn.*, vol. 2, no. 2, pp. 10–15, Jul. 1954.
[6] Y. Serizawa, Y. Sekimoto, M. Kamikura, W. Shan, and T. Ito, "A 400–500 GHz balanced SIS mixer with a waveguide quadrature hybrid coupler," *Int. J. Infrared Millim. Waves*, vol. 29, no. 9, pp. 846–861, June 2008.
[7] A. R. Kerr, "On the noise properties of balanced amplifiers," ALMA, , Memo 227, Sep. 1998.
[8] S. A. Maas, "Microwave mixers," 2nd ed. NRAO, Charlottesville, VA.
[9] E. J. Wilkinson, "An N-way hybrid power divider," *IRE, Microw. Theory Techn.*, vol. MTT-13, pp. 116–118, Jan. 1960.
[10] M. P. Westig, K. Jacobs, J. Stutzki, M. Schultz, M. Justen, and C. E. Honingh, "Balanced superconductor-insulator-superconductor mixer on a 9 $\mu$m silicon membrane," *Supercond. Sci. Technol.*, vol. 24, no. 8, p. 6, Aug. 2011.
[11] Virginia Diodes Inc., Charlottesville, VA, "Virginia Diodes, Inc., homepage," [Online]. Available: http://vadiodes.com/
[12] Precision Cryogenics Systems Inc., Indianapolis, IN, "Cryogenics Systems Inc., homepage," [Online]. Available: http://www.precisioncryo.com/
[13] S. M. X. Claude and C. T. Cunningham, "Design of a sideband-separating balanced SIS mixer based on waveguide hybrids," Herzberg Inst. Astrophys., Canada, ALMA Memo 316, Sep. 2000.
[14] Ansoft Corporation, Pittsburgh, PA.
[15] Custom Microwave Inc., Longmont, CO.
[16] American Technical Ceramics, Huntington Station, NY.
[17] W. Menzel, L. Zhu, K. Wu, and F. Bögelsack, "On the design of novel compact broadband planar filters," *IEEE Trans. Microw. Theory Tech.*, vol. 51, no. 2, pp. 364–370, Feb. 2003.
[18] B. Bumble, "Private communication," Jet Propulsion Laboratory (JPL), Pasadena, CA, 2003.
[19] J. Ward, F. Rice, and J. Zmuidzinas, University of Virginia, Charlottesville, VA, "Supermix: A flexible software library for high-frequency circuit simulation, including SIS mixers and superconducting components," in *Proc. 10th Int. Symp. Space Terahertz Tech.*, 1999, pp. 269–281.
[20] Sonnet, Liverpool, NY, "Sonnet software," 2005. [Online]. Available: http://www.sonnetusa.com/
[21] Anritsu Company, Morgan Hill, CA, Model: MG3690B, 2008. [Online]. Available: http://www.anritsu.com/en-US/Products-Solutions/Test-Measurement/RF-Microwave/Signal-Generators
[22] Micro-coax Inc., Pottstown, PA, UFB142A. [Online]. Available: http://www.micro-coax.com
[23] Norden Millimeter Inc., Placerville, CA, Model: N09-2414. [Online]. Available: http://www.nordengroup.com/index.html
[24] Micro Lambda Wireless Inc., Fremont, CA, Model: MLFP-41840RS. [Online]. Available: http://www.microlambdawireless.com
[25] Ditom Microwave Inc., Fresno, CA, Model: D3I2004. [Online]. Available: http://www.ditom.com
[26] Marki Microwave Inc., Morgan Hill, CA, Model: FX0069. [Online]. Available: http://www.markimicrowave.com
[27] Pacific Millimeter Products, Inc., Golden, CO, Models: E3, E3+, W3, W3-. [Online]. Available: http://pacificmillimeterproducts.lbu.com/index.html
[28] Zen Machine and Scientific Instruments, Lyons, CO..
[29] Micropac Industries Inc., Garland, TX, Model: 52416. [Online]. Available: http://www.micropac.com
[30] Teledyne Microwave, Mountain View, CA, "Microwave YIG filters," Application Note OFOISR App 06-S-1942.

[31] R. Chamberlin and J. W. Kooi, "CSO YIG filter tuning performance," Internal Report, Apr. 2011.
[32] H. Wang, L. Samoska, T. Gaier, A. Peralta, H.-H. Liao, Y. C. Leong, S. Weinreb, Y. C. Chen, M. Nishimoto, and R. Lai, "Power-amplifier modules covering 70–113 GHz using MMICs," *IEEE Trans. Microw. Theory Tech.*, vol. 49, no. 1, pp. 9–16, Jan. 2001.
[33] Th. de Graauw, F. P. Helmich, T. G. Phillips, J. Stutzki, E. Caux, N. D. Whyborn, P. Dieleman, P. R. Roelfsema, H. Aarts, R. Assendorp, R. Bachiller, W. Baechtold, A. Barcia, D. A. Beintema, V. Belitsky, A. O. Benz, R. Bieber, A. Boogert, C. Borys, B. Bumble, P. Caïs, M. Caris, P. Cerulli-Irelli, G. Chattopadhyay, S. Cherednichenko, M. Ciechanowicz, O. Coeur-Joly, C. Comito, A. Cros, A. de Jonge, G. de Lange, B. Delforges, Y. Delorme, T. den Boggende, J.-M. Desbat, C. Diez-González, A. M. Di Giorgio, L. Dubbeldam, K. Edwards, M. Eggens, N. Erickson, J. Evers, M. Fich, T. Finn, B. Franke, T. Gaier, C. Gal, J. R. Gao, J.-D. Gallego, S. Gauffre, J. J. Gill, S. Glenz, H. Golstein, H. Goulooze, T. Gunsing, R. Güsten, P. Hartogh, W. A. Hatch, R. Higgins, E. C. Honingh, R. Huisman, B. D. Jackson, H. Jacobs, K. Jacobs, C. Jarchow, H. Javadi, W. Jellema, M. Justen, A. Karpov, C. Kasemann, J. Kawamura, G. Keizer, D. Kester, T. M. Klapwijk, Th. Klein, E. Kollberg, J. Kooi, P.-P. Kooiman, B. Kopf, M. Krause, J.-M. Krieg, C. Kramer, B. Kruizenga, T. Kuhn, W. Laauwen, R. Lai, B. Larsson, H. G. Leduc, C. Leinz, R. H. Lin, R. Liseau, G. S. Liu, A. Loose, I. López-Fernandez, S. Lord, W. Luinge, A. Marston, J. Martín-Pintado, A. Maestrini, F. W. Maiwald, C. McCoey, I. Mehdi, A. Megej, M. Melchior, L. Meinsma, H. Merkel, M. Michalska, C. Monstein, D. Moratschke, P. Morris, H. Muller, J. A. Murphy, A. Naber, E. Natale, W. Nowosielski, F. Nuzzolo, M. Olberg, M. Olbrich, R. Orfei, P. Orleanski, V. Ossenkopf, T. Peacock, J. C. Pearson, I. Peron, S. Phillip-May, L. Piazzo, P. Planesas, M. Rataj, L. Ravera, C. Risacher, M. Salez, L. A. Samoska, P. Saraceno, R. Schieder, E. Schlecht, F. Schlöder, F. Schmülling, M. Schultz, K. Schuster, O. Siebertz, H. Smit, R. Szczerba, R. Shipman, E. Steinmetz, J. A. Stern, M. Stokroos, R. Teipen, D. Teyssier, T. Tils, N. Trappe, C. van Baaren, B.-J. van Leeuwen, H. van de Stadt, H. Visser, K. J. Wildeman, C. K. Wafelbakker, J. S. Ward, P. Wesselius, W. Wild, S. Wulff, H.-J. Wunsch, X. Tielens, P. Zaal, H. Zirath, J. Zmuidzinas, and F. Zwart, "The Herschel-heterodyne instrument for the far-infrared (HIFI)," *Astronomy Astrophys.*, vol. 518, p. L6, 2010.
[34] G. L. Pilbratt, J. R. Riedinger, T. Passvogel, G. Crone, D. Doyle, U. Gageur, A. M. Heras, C. Jewell, L. Metcalfe, S. Ott, and M. Schmidt, "Herschel space observatory. An ESA facility for far-infrared and submillimetre astronomy," *Astronomy Astrophys.*, vol. 518, p. L1, 2010.
[35] A. R. Kerr, H. Moseley, E. Wollack, W. Grammer, G. Reiland, R. Henry, and K. P. Stewart, "MF-112 and MF-116: Compact wavguide loads and FTS measurements at room temperature and 5 K," NRAO, ALMA, Charlotteville, VA, Memo 494, Mar. 14, 2004.
[36] Millitech Inc., Northhampton, MA. [Online]. Available: http://www.millitech.com
[37] J. W. Kooi, G. Chattopadhyay, M. Thielman, T. G. Phillips, and R. Schieder, "Noise stability of SIS receivers," *Int. J. Infrared Millim. Waves*, vol. 21, no. 5, pp. 689–716, May 2000.
[38] R. Schieder and C. Kramer, "Optimization of heterodyne observations using Allan variance measurements," *Astron. Astrophys.*, vol. 373, pp. 746–756, Jul. 2001.
[39] J. W. Kooi, J. J. A. Baselmans, A. Baryshev, R. Schieder, M. Hajenius, J. R. Gao, T. M. Klapwijk, B. Voronov, and G. Gol'tsman, "Stability of heterodyne terahertz receivers," *J. Appl. Phys.*, vol. 100, p. 064904, Sep. 2006.
[40] P. F. Goldsmith, *QuasiOptical Systems*. Piscataway, NJ: IEEE Press, 1998.
[41] R. Hesper, B. D. Jackson, A. M. Baryshev, J. Adema, K. Wielinga, M. Kroug, T. Zijlstra, G. Gerlofsma, M. Bekema, K. Keizer, H. Schaeffer, J. Barkhof, F. P. Mena, A. Koops, R. Rivas, T. M. Klapwijk, and W. Wild, "Design and development of a 600–720 GHz receiver cartridge for ALMA band 9," in *Proc. 16th Int. Symp. Space Terahertz Tech.*, 2005, p. 110.
[42] N. Erickson, Harvard Univ., "A fast, very sensitive calorimetric power meter for millimeter to submillimeter wavelengths," in *Proc. 13th Int. Symp. Space Terahertz Tech.*, 2002, pp. 301–307.
[43] A. J. Gatesman, J. Waldman, M. Ji, C. Musante, and S. Yagvesson, "An antireflection coating for silicon optics at terahertz frequencies," *IEEE Microw. Wirel. Compon. Lett.*, vol. 10, no. 7, pp. 264–266, Jul. 2000.
[44] Saint-Gobain Performance Plastics, Global, "Saint-Gobain performance plastics," [Online]. Available: http://www.norton-films.com/zitexg-filter-membranes.aspx
[45] Carlisle Plastics Company, Inc., New Carlisle, OH.
[46] D. Koller, A. R. Kerr, and G. A. Ediss, "Proposed quartz vacuum window designs for ALMA bands 3–10," NRAO, Charlottesville, VA, 2001. [Online]. Available: http://www.alma.nrao.edu/memos/html-memos/alma397/memo397.pdf
[47] D. Maier, A. Barbier, B. Lazareff, and K. F. Schuster, "The ALMA band 7 mixer," presented at the 16th Int. Symp. Space Terahertz Tech., Chalmers, Göteborg, Sweden, 2005.

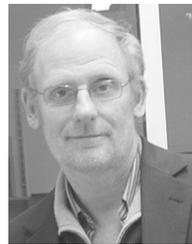

**Jacob W. Kooi** was born in Geldrop, The Netherlands, on July 12, 1960. He received the B.S. degree in microwave engineering from the California Polytechnic State University, San Luis Obispo, CA, in 1985, the M.S. degree in electrical engineering from the California Institute of Technology, Pasadena, in 1992, and the Ph.D. degree in physics from the Rijksuniversity Groningen, The Netherlands, in 2008.

His research interests include the areas of millimeter and submillimeter wave technology, low noise amplifiers, multipliers, instrumental stability, Fourier optics, and their application to astronomy.

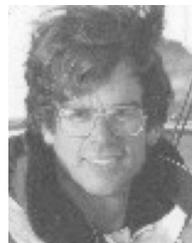

**Richard Chamberlin** received the B.S. degree from the University of California, Santa Barbara, in 1984, and the Ph.D. degree (under George B. Benedek) from the Massachusetts Institute of Technology, Cambridge, in 1991, both in physics.

He served in the United States Air Force from 1975 to 1979 as a Weather Observer. In 1995, he was the first winter-over scientist with the pioneering Antarctic Submillimeter Telescope and Remote Observatory which he helped design, build, and test while at Boston University. From 1996 to 2010, he was the Technical Manager of the Caltech Submillimeter Observatory located on the summit of Mauna Kea on the Island of Hawaii. Since 2010 he is affiliated one half time with California Institute of Technology, Pasadena, CA, and one half time with the University of Colorado and the National Institute of Standards and Technology both in Boulder, CO. His research interests include THz instrumentation and atmospheric science.

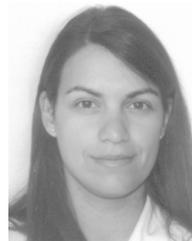

**Raquel R. Monje** received the M.S. degree in telecommunication engineering from Universidad Europea de Madrid, Madrid, Spain, in 2003, the M.S. degree in digital communications system and technology and the Ph.D. degree in radio and space science from Chalmers University of Technology, Gothenburg, Sweden, in 2004 and 2008, respectively. Her Ph.D. thesis was on low noise superconductor-insulator-superconductor (SIS) mixers for submillimeter and millimeter-wave astronomy.

She is currently a Senior Postdoctoral Scholar with California Institute of Technology, Pasadena. Her research interests include microwave technology, SIS mixers, millimeter and submillimeter wave heterodyne receivers for astronomy and the associated science resulting from observations.

**Brian Force** is an RF/microwave Engineer with Caltech Submillimeter Observatory located on the summit of Mauna Kea on the Island of Hawaii. As part of his technical duties he supports the telescope and helps observers operate it.

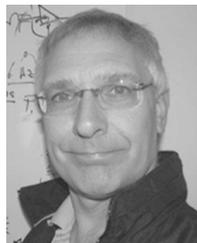

**David Miller** received the B.S. and M.S. degrees in electrical engineering from the California State Polytechnic University, Pomona, in 1990 and 1998, respectively. He is currently pursuing the M.A. degree in theology from Talbot School of Theology, La Mirada, CA.

His research interests include microwave engineering, low-noise and high-stability electronics, and the design, construction, and testing of submillimeter receivers for airborne and land-based observatories. The astronomical receivers are based on either the heterodyne technique, such as SIS and HEB mixers, or the direct detection technique, such as MKIDs and bolometers.

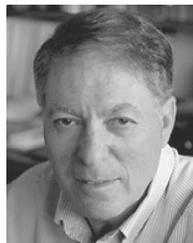

**Tom G. Phillips** received the B.A., M.A., and D.Phil. degrees from Oxford University, Oxford, U.K..

His graduate studies were in low-temperature physics. After one year at Stanford University, he returned to Oxford for two years and then moved to the Bell Laboratories Physics Research Laboratory, Murray Hill, NJ. There he developed techniques for millimeter and submillimeter wave detection for astronomy. In 1975, he spent one year at London University as University Reader in physics. In 1980, he joined the faculty of the California Institute of Technology (Caltech) as Professor of Physics. At Caltech, he took on the task of construction of the Owens Valley Radio Observatory Millimeter Wave Interferometer, as Associate Director of the Observatory. In 1982, he became Director Designate for the Caltech Submillimeter Wave Observatory, to be constructed in Hawaii, and in 1986, on successful completion of the construction, became Director. His current research interests include molecular and atomic spectroscopy of the interstellar medium and in the development of superconducting devices for submillimeter-wave detection.